\documentclass[a4paper,11pt]{article}

\usepackage{contribution}

\newcommand{\weblink}[2][]{%
    \ifthenelse{\equal{#1}{}}%
    {\textnormal{\url{#2}}}%
    {\textnormal{\href{#2}{#1}}}%
}

\def\beq{\begin{equation}}
\def\eeq#1{\label{#1}\end{equation}}
\def\eeqn{\end{equation}}

\def\beqa{\begin{eqnarray}}
\def\eeqa#1{\label{#1}\end{eqnarray}}
\def\eeqan{\end{eqnarray}}

\let\bar=\overbar

\def\Dslash{\not{\hbox{\kern-4pt $D$}}}
\def\dslash{\not{\hbox{\kern-2pt $\del$}}}

\def\msb{{\bar{\ssstyle M \kern -1pt S}}}

\newcommand{\contribution}[7][]{%
  \clearpage
  \thispagestyle{plain}
  \ifthenelse{\equal{#1}{}}
  {\hypersetup{pdftitle={#2}}}
  {\hypersetup{pdftitle={#1}}}
  \hypersetup{pdfauthor={{#3} {#4}}}
  {\centering\normalfont\LARGE\bfseries\sffamily #2 \par\nobreak}
  \lhead{}
  \chead{%
    \textit{\footnotesize XIV International Conference on Hadron Spectroscopy
      (\weblink[\textit{hadron2011}]{http://www.hadron2011.de}), 13-17 June 2011, Munich, Germany}%
  }
  \rhead{}
  \bigskip
  \begin{center}
    {#3} {#4}\ifthenelse{\equal{#6}{}}{}{\footnote{\weblink[#6]{mailto:#6}}}
    \ifthenelse{\equal{#7}{}}{}{#7} \\
    \textit{#5}
  \end{center}
  \bigskip
}

\renewcommand{\abstract}[1]{%
  \begin{center}
    \begin{minipage}{0.85\textwidth}
      \begin{footnotesize}
        #1
      \end{footnotesize}
    \end{minipage}
  \end{center}
  \bigskip
}

\begin{document}

%
%
%
%
%
{  

\newcommand\TT{\rule{0pt}{2.5ex}}        
\newcommand\BB{\rule[-1.0ex]{0pt}{0pt}}  

\newcommand{\be}{\begin{equation}}
\newcommand{\en}{\end{equation}}
\newcommand{\bea}{\begin{eqnarray}}
\newcommand{\ena}{\end{eqnarray}}
\newcommand{\lbl}[1]{\label{eq:#1}}
\newcommand{\lbltab}[1]{\label{tab:#1}}
\newcommand{\lblfig}[1]{\label{fig:#1}}
\newcommand{\lblsec}[1]{\label{sec:#1}}
\newcommand{\rf}[1]{(\ref{eq:#1})}
\newcommand{\Table}[1]{\ref{tab:#1}}
\newcommand{\fig}[1]{\ref{fig:#1}}
\newcommand{\sect}[1]{\ref{sec:#1}}
\newcommand{\braque}[1]{{\langle #1 \rangle}}
\newcommand{\bc}{\begin{center}}
\newcommand{\ec}{\end{center}}
\newcommand{\bt}{\begin{tabular}}
\newcommand{\et}{\end{tabular}}
\newcommand{\ba}{\begin{array}}
\newcommand{\ea}{\end{array}}
\newcommand{\gapprox}{%
\mathrel{%
\setbox0=\hbox{$>$}\raise0.3ex\copy0\kern-\wd0\lower0.80ex\hbox{$\sim$}}}

\newcommand{\lapprox}{%
\mathrel{%
\setbox0=\hbox{$<$}\raise0.6ex\copy0\kern-\wd0\lower0.65ex\hbox{$\sim$}}}
\newcommand{\inleft}{%
\mathrel{%
\setbox0=\hbox{$<$}\copy0\kern-0.5\wd0\lower1.1\ht0\hbox{$\scriptstyle{in}$}}}
\newcommand{\inright}{%
\mathrel{%
\setbox0=\hbox{$>$}\copy0\kern-0.5\wd0\lower1.1\ht0\hbox{$\scriptstyle{in}$}}}
\newcommand{\outleft}{%
\mathrel{%
\setbox0=\hbox{$<$}\copy0\kern-0.5\wd0\lower1.1\ht0\hbox{$\scriptstyle{out}$}}}
\newcommand{\outright}{%
\mathrel{%
\setbox0=\hbox{$>$}\copy0\kern-0.5\wd0\lower1.1\ht0\hbox{$\scriptstyle{out}$}}}
\newcommand{\e}{{\rm e}}
\newcommand{\im}{{\rm Im\,}}
\newcommand{\re}{{\rm Re\,}}
\newcommand{\Kbar}{\bar{K}}
\newcommand{\mpi }{m_\pi}
\newcommand{\mpid}{m_\pi^2}
\newcommand{\mpic}{m_\pi^3}
\newcommand{\fpid}{F_\pi^2}
\newcommand{\mpiq}{m_\pi^4}
\newcommand{\mkd}{m_K^2}
\newcommand{\mbkd}{\bar{m}_K^2}
\newcommand{\mkq}{m_K^4}
%

\contribution[]  
{Properties of light scalar mesons in the complex plane}  
{Bachir}{Moussallam}  
{Groupe de physique th\'eorique, IPN, Universit\'e Paris-sud 11, 91406
Orsay, France}  
{}  
{}  
%

\abstract{%
The flavour and glue structure of the light
scalar mesons in QCD are probed by studying the couplings of the
$\sigma(600)$ and $f_0(980)$ to the operators $\bar{q}q$,
$\theta_\mu^\mu$ and to two photons. 
}
%

\section{Introduction}
A precise identification of the exotic contents of the light scalar
mesons is an old challenge for hadron physics.
The idea of using the Roy dispersive representations in the complex
plane~\cite{CCL} decisively clarified the status of the broad $\sigma$
resonance and led to an accurate determination of its mass and width. 
The $I=0$ scalars have the same quantum numbers as the expected
lightest glueball. If all quark masses were heavy in QCD
($m_q\gapprox 1 GeV$) the glueball would be an extremely narrow meson
(decaying only to two photons) with a mass slightly below two
GeV~\cite{Morningstar}. Properties of the glueball in physical QCD,
i.e. with three light quarks, are not yet computable in lattice simulations.
A scenario suggested in~\cite{narisonveneziano} is that there could be
two glueballs below 2 GeV, the lightest one could be the $\sigma$. 
Another question concerns the identification of a flavour multiplet: do the
$\sigma$, $\kappa$, $a_0(980)$, $f_0(980)$ mesons belong to a nonet?
This is far from obvious, e.g. several properties of the
$a_0(980)$, $f_0(980)$ seem well explained by a $K\Kbar$ molecule
model~\cite{Weinstein:1990gu}. 
I discuss below the evaluation of couplings of the  $\sigma$ and
$f_0(980)$ to operators $j_S$ using complex plane methods, which
should provide quantitative inputs for answering some of these questions
on the light scalars. Taking $j_S=\theta_\mu^\mu$ probes the gluon
content, while $j_S=\bar{u}u+\bar{d}d,\ \bar{s}s$ probes
the quark-antiquark content.  The couplings to two photons will also
be updated.

\section{Poles and residues from an extended $S$-wave Roy solution:}
Pion-pion partial-wave amplitudes obey a set of coupled integral
representations as a consequence of analyticity and crossing
symmetry~\cite{Roy}. On the real axis, at low energy, they
combine with the unitarity relation into a set of non-linear equations
which strongly constrain the $S$ and the $P$-wave
amplitudes. These equations were reconsidered recently in great
detail~\cite{ACGL}. Solutions were generated for given  scattering
lengths $a_0^0$, $a_0^2$ with a matching point
$\sqrt{s_A}=0.8$ GeV (i.e. using experimental inputs for $s > s_A$). 
The new experimental results on $K_{l4}$ decays, analyzed
with these solutions lead to very precise determinations of the
scattering lengths~\cite{lastNA48} 
\bea
&& a_0^0=0.2196\pm0.0028_{\hbox{stat}}\pm0.0020_{\hbox{sys}},\\
&& a_0^2=-0.0444\pm0.0007_{\hbox{stat}}\pm 0.0005_{\hbox{sys}}\pm0.0008_{\hbox{ChPT}}
\nonumber
\ena
\begin{floatingtable}[r]{
\bt{c|cccc}\hline\hline
\TT\BB $\eta_0^0$ & $\delta_A$ & $\delta_K$ & $\hat{\chi}^2_{\hbox{Hyams}}$ &
$\hat{\chi}^2_{\hbox{Kaminski}}$\\ \hline
shallow & $\left(80.9\pm 1.4\right)^\circ$ & 
$\left(190^{+5}_{-10}\right)^\circ$ &2.7   &  1.9 \\  
deep & $\left(82.9\pm 1.7\right)^\circ$   & 
$\left(200^{+5}_{-10}\right)^\circ$ &2.2   &  1.3\\ \hline\hline
\et}
\caption{ Fitted values of  $\delta_A$ and $\delta_K$
  corresponding to two different
  central values (shallow-dip, deep-dip) for $\eta_0^0$.}
\lbltab{fitres}
\end{floatingtable}
In order to improve the constraints on the $f_0(980)$ resonance, we
begin by constructing a solution for the $I=0$ $S$-wave over an
extended region with a matching point $s_K=4\mkd$. Increasing the
matching point from $s_A$ to $s_K$ leads to two differences: 
a) the matching point coincides with a two-particle threshold, the
phase-shift has a cusp and one cannot use the continuity of the
derivative argument to eliminate unphysical solutions, 
b) the multiplicity of the solutions (which depends on the
phase-shift at the matching point~\cite{pomponiuwanders}) increases by one
unit. It turns out that unphysical solutions can still be identified
provided the phase-shift is not too large ($\delta_K=\delta_0^0(s_K)\lapprox
225^\circ$). The solutions then depend on one arbitrary parameter
which can be taken as the value of the phase at one energy, e.g.
$\delta_A=\delta_0^0(s_A)$. It is then determined by
fitting the experimental phase-shifts in the range $[s_A,s_K]$ with
the Roy solutions. One, in fact, can fit not only $\delta_A$ but also 
$\delta_K$ which is not very accurately determined from the data above $s_K$.

\begin{floatingtable}[h]{
\bt{c|cc}\hline\hline
\TT\BB & $\sqrt{z_S}$ (MeV) & $\dot{S}_0^0(z_S)$ (GeV$^{-2}$) \\ \hline
\TT\BB $\sigma(600)$ & $\left(442^{+5}_{-8}\right) -i\left(274^{+6}_{-5}\right)$ &  
$-\left(0.75^{+0.10}_{-0.15}\right)-i\left(2.20^{+0.14}_{-0.10}\right)$  \\
\TT\BB $f_0(980)$    & $\left(996^{+4}_{-14}\right) -i\left(24^{+11}_{-3}\right)$ & 
$-\left(1.1^{+3.0}_{-0.4}\right)-i\left(6.6^{+0.8}_{-1.0}\right)$   \\ \hline\hline
\et}
\caption{Poles  and $S$-matrix derivatives  from  the
  extended Roy solution. }
\lbltab{cxpoles}
\end{floatingtable}
The shape of the inelasticity, which sets in effectively at $s \ge s_K$
affects the phase-shift below $s_K$ via the Roy
equation. The central value of $\eta_0^0$ fitted to the determination
from $\pi\pi$ production experiments displays a deep dip
structure~\cite{hyams73}. The one based on the
inelastic channels displays a shallower dip (while compatible within
the errors).  
The recent analysis of ref.~\cite{GKPY3}, using constraints from a
variant of the Roy equations in the region $\sqrt{s}\le 1.1$ GeV,
favours a deep dip structure. Our fit leads to the same conclusion, see
table~\Table{fitres}.

Next, one can compute the amplitude  for complex energies. 
The position of the poles on the second Riemann sheet correspond
to the zeros of the $S$-matrix $S_0^{0}(z)=1-2\sigma^\pi(z)
t_0^0(z)$ (see e.g.~\cite{CCL}) with
$\sigma^\pi(z)=\sqrt{4\mpid/z-1}$. The results for the zeros and for
the corresponding derivatives of $S_0^0$  (needed for the
determination of the residues) are shown in table~\Table{cxpoles}. 

\section{Couplings to two photons}
Couplings of the light scalars to two
photons can be determined from the amplitudes $\gamma\gamma\to
\pi^0\pi^0, \pi^+\pi^-$. As a consequence of analyticity and unitarity, the
partial-wave amplitudes $h^I_{J,\lambda\lambda'}(s)$ satisfy
Omn\`es-type dispersive representations. A two-channel
representation for $h^0_{0,++}(s)$ was reconsidered
recently~\cite{garciamartinmou}.  It should be valid in a range 
$\sqrt{s}\lapprox 1.1-1.2$ GeV where it is a good approximation to retain
just two channels ($\pi\pi$,  $K\Kbar$) in the unitarity relation and
has the form 
\bea\lbl{2chanrepres}
&& \left(
\ba{l} 
h^0_{0,++}(s)\\
k^0_{0,++}(s)
\ea\right) =
\left(
\ba{l} 
\bar{h}^{0,Born}_{0,++}(s)\\
\bar{k}^{0,Born}_{0,++}(s)
\ea\right) + 
\overline{\overline{\Omega}}(s)\times\Bigg[
\left(\ba{l} 
b^{(0)} s +b^{'(0)} s^2\\
b_K^{(0)} s +b_K^{'(0)} s^2
\ea\right) 
+{s^3\over\pi}\int_{-\infty}^{-s_0}{ds'\over(s')^3(s'-s)}\times 
\\
&&\quad \overline{\overline{\Omega}}^{-1}(s')\im
\left(\begin{array}{l} 
\bar{h}^{0,Res}_{0,++}(s')\\
\bar{k}^{0,Res}_{0,++}(s')
\end{array}\right)
-{s^3\over\pi}
\int_{4\mpid}^\infty 
{ds'\over (s')^3(s'-s)} \im \overline{\overline{\Omega}}^{-1}(s')
\left(\begin{array}{l} 
\bar{h}^{0,Born}_{0,++}(s')\\
\bar{k}^{0,Born}_{0,++}(s')
\end{array}\right)
\Bigg]\ .\nonumber
\ena
The key ingredient in this equation is 
the  Omn\`es matrix $\overline{\overline{\Omega}}$
which must be computed numerically from the $T$-matrix. 
Eq.~\rf{2chanrepres} involves integrals over the left-hand cut
of the amplitude which is generated by cross-channel
singularities:  the pion pole ($\bar{h}^{I,Born}_{0,++}$) and
multi-pion cuts which are approximated by resonance poles. 
It also involves four polynomial parameters which 
appear upon writing over-subtracted dispersion relations such
as to cutoff  contributions from higher energy regions.
They  were determined in
ref.~\cite{garciamartinmou} from a chirally constrained fit of the
Belle data~\cite{Belle1Belle2}. The representation
~\rf{2chanrepres} then allows one to compute the  amplitude
$h^0_{0,++}(s)$ for complex values of $s$. Extension to the second
Riemann sheet is performed from standard formulas~\cite{pennington06}
involving $S_0^0(s)$. The couplings $g_{S\pi\pi}$ and
$g_{S\gamma\gamma}$ are identified from the residues and one then
defines the $2\gamma$ decay width~\cite{pennington06} in terms of
$\vert g_{S\gamma\gamma}\vert$. We find the following results for the $\sigma$
and $f_0(980)$ (in KeV)  
\be\lbl{w2gamma}
\Gamma_{\sigma(600)\to 2\gamma}=\left(2.08\pm 0.20\,
^{+0.07}_{-0.04}\right) ,\ 
\Gamma_{f_0(980)\to 2\gamma}  =  \left(0.29 \pm0.21 \,
^{+0.02}_{-0.07}\right) \ .
\en
The result for the $\sigma$ agrees with that presented
by Hoferichter at this conference and is somewhat smaller than the one
in ref.~\cite{pennington06}.
\section{Couplings to gluonic and quark-antiquark operators}
Let us introduce, formally at first, couplings of the scalar mesons to
$\bar{q}q$ operators and to the trace of the energy-momentum tensor
$\theta_\mu^\mu$, 
\be\lbl{couplings}
\braque{0\vert \bar{u}u+  \bar{d}d \vert S}  =\sqrt2 B_0 \,C_S^{uu},\quad
\braque{0\vert \bar{s}s \vert S} = B_0 \,C_S^{ss} ,\quad
\braque{0\vert \theta_\mu^\mu \vert S}  = m_S^2 \,C_S^{\theta}\ .
\en
In order to give  precise meaning to these one considers the correlators  
\be
\Pi_{jj}(p^2)=i\int d^4x\, e^{ipx} \braque{0\vert T j_S(x) j_S(0)\vert 0}
\en
where $j_S(x)$ is one of the scalar operators considered above. 
The discontinuity along the cut in the elastic region
$4\mpid\le s\le 16\mpid$  follows from the
K\"allen-Lehmann representation and allows one to deduce the
expression for the second sheet extension of the correlator,
\be
\Pi^{II}_{jj}(z)= \Pi_{jj}(z)+ {3\over 16\pi} {\sigma^\pi(z) \left(
  F_j(z)\right)^2\over  1- 2\sigma^\pi(z) t_0^0(z)}\ .
\en
$F_j(s)$ is the pion form factor associated with
$j_S$. $\Pi^{II}_{jj}(z)$ exhibits poles  at $z=z_S$ corresponding to the
scalar mesons. The coupling constants~\rf{couplings} are naturally
identified from the residues and are thus determined in the terms of
the value of the form-factor $F_j(z_S)$.

\begin{floatingtable}[h]{
\bt{c|cc}\hline\hline
\TT  & $\sigma(600)$  & $f_0(980)$   \\ \hline
\TT$\vert C^{uu}_S\vert$ & $206\pm4 ^{+4}_{-6}$  & $82\pm31 ^{+12}_{-7}$ \\
\TT$\vert C^{ss}_S\vert$ & $17\pm5  ^{+1}_{-7}$  & $146\pm44^{+14}_{-7}$\\ 
\TT$\vert C^{\theta}_S\vert$ & $197\pm15^{+21}_{-6}$& $114\pm44 ^{+22}_{-7}$ \\
\hline\hline
\et}
\caption{\sl Absolute values (in MeV) of the  couplings of the
  $\sigma$ and $f_0(980)$  to  operators. } 
\lbltab{qqbarcoupl}
\end{floatingtable}
Analyticity properties of the form-factors allows them to
be expressed in terms of the Omn\`es matrix. For this purpose,
one considers the two component vector $\bar{F}(s)$ such that
$\bar{F}_1(s)=F^\pi_j(s)$, $\bar{F}_2(s)=2/\sqrt3 F_j^K(s)$. Multiplying
$\bar{F}(s)$ by the inverse of the Omn\`es matrix removes the
right-hand cut up to a point $s_2$ where two-channel unitarity is no
longer a good approximation. One then deduces that in
the region $\vert s\vert \lapprox s_2$ the form-factors obey the
following representation,
\be\lbl{omffactor}
\left(\ba{r}
F_j^\pi(s)\\[0.1cm]
{2\over\sqrt3} F_j^K(s)
\ea\right)= 
\left(\ba{cc}
\Omega_{11}(s) & \Omega_{12}(s)\\[0.1cm]
\Omega_{21}(s) & \Omega_{22}(s)
\ea\right)\left(\ba{c}
\alpha +\alpha' s\\
\beta  +\beta' s
\ea\right)\ .
\en
The polynomial parameters in eq.~\rf{omffactor} can be estimated from the
chiral expansion at leading order~\cite{DGL90}. The magnitudes of the
resulting couplings  are shown in
table~\Table{qqbarcoupl}: the first and second error  in the entries
reflect the uncertainties in the polynomial parameters
and in the Omn\`es matrix respectively  . 

In summary, one finds that both
the $\sigma$ and the $f_0(980)$ couple significantly to
$\theta_\mu^\mu$. The sum of the couplings is in qualitative agreement
with the result of ref.~\cite{narisonveneziano}. The coupling of the
$\sigma$ to the $\bar{u}u+\bar{d}d$ operator can be compared to the
coupling of the $a_0(980)$ to $\bar{u}d$. Converting the finite-energy
sum rule result of Maltman~\cite{Maltman:1999jn} to the present normalisation
gives $\vert C_{a_0(980)}^{ud}\vert =197\pm37$ MeV. Thus, the $\sigma$
and $a_0$ couplings are nearly equal, which is  compatible with an
assignment of both mesons into the same multiplet. We have also
estimated the $\kappa$ meson coupling $\vert C_{\kappa(800)}^{us}\vert
\simeq 156$ MeV. The couplings of the light
scalars $\sigma$, $\kappa$, $a_0(980)$, $f_0(980)$ to $\bar{q}q$
operators are therefore not unusually
small, as one could expect in a naive tetraquark model, and their values
are compatible with a nonet assignment.

%

}  


\end{document}